\documentclass[twocolumn,showpacs,preprintnumbers,amsmath,amssymb]{revtex4}

\usepackage{graphicx}
\usepackage{dcolumn}
\usepackage{bm}

\begin{document}


\title{\large Coupling symmetry of quantum dot states}

\author{M. C. Rogge$^1$}
\email{rogge@nano.uni-hannover.de}
\author{B. Harke$^1$}
\author{C. Fricke$^1$}
\author{F. Hohls$^{1,2}$}
\author{M. Reinwald$^3$}
\author{W. Wegscheider$^3$}
\author{R.~J. Haug$^1$}
 \affiliation{$^1$Institut f\"ur Festk\"orperphysik, Universit\"at Hannover,
  Appelstra\ss{}e 2, D-30167 Hannover, Germany \\ $^2$Cavendish Laboratory, University of Cambridge,
  Madingley Road, Cambridge CB3 OHE, Great Britain \\
$^3$Angewandte und Experimentelle Physik, Universit\"at
Regensburg, D-93040 Regensburg, Germany}

\date{\today}

\begin{abstract}
With non-invasive methods, we investigate ground and excited
states of a lateral quantum dot. Charge detection via a quantum
point contact is used to map the dot dynamics in a regime where
the current through the dot is too low for transport measurements.
In this way we investigate and compare the tunneling rates from
the dot to source and drain. We find a symmetry line on which the
tunneling rates to both leads are equal. In this situation ground
states as well as excited states influence the mean charge of the
dot. A detailed study in this regime reveals that the coupling
symmetry depends on the number of states contributing to transport
and on the spatial distribution of individual states.
\end{abstract}

\maketitle

Transport measurements on 2D, 1D and 0D \cite{Kouwenhoven-97}
systems have provided important insights into semiconductor-based
nanophysics during past decades. In recent years non-invasive
probing methods have complemented the researchers' utilities.
These methods are based on charge detection via capacitive
coupling between adjacent nanostructures. A reliable way to probe
quantum dots (QDs) is to use 1D quantum point contacts (QPCs) as
charge detectors. This was successfully demonstrated on lateral
single and double dots fabricated using electron beam lithography
\cite{Field-93,Elzerman-03} or local anodic oxidation with atomic
force microscopes \cite{Nemutudi-04,Schleser-APL04,Fricke-05}. In
pulsed measurements QPCs can also be used for spin readout
\cite{Elzerman-Nature04} or to probe excited dot states
\cite{Elzerman-APL04,Johnson-PRB05}.


In this Letter we demonstrate that one can probe excited dot
states in dc measurements. We study the tunneling rates $\Gamma_S$
and $\Gamma_D$ for electrons traversing the dot via source and
drain leads. Indeed, these rates can be measured individually when
only one barrier is used while the other one is kept far open
without a QD. But these results cannot be transferred one-to-one
to a regime with a QD as the wavefunction overlap is changed when
defining the dot. Thus this method gives only a rough estimate.
Instead we use the QPC to investigate the rates in a very
sensitive way keeping the dot in the Coulomb blockade regime.
Configurations can be detected where both rates are equal. The
positions of these configurations in the parameter space roughly
follow a line. Detailed analysis reveals that the coupling
symmetry depends on individual QD states and on the number of
channels used for transport. With these insights, asymmetry can be
introduced systematically, which is useful for the investigation
of effects that arise from the interplay of dot and leads (e.g.,
Kondo effect, Fano effect, spin blocking mechanisms or
conductances of ground and excited states). Asymmetric barriers
can, for example, lead to negative differential conductance
\cite{Cavaliere-04}.

\begin{figure}
\includegraphics{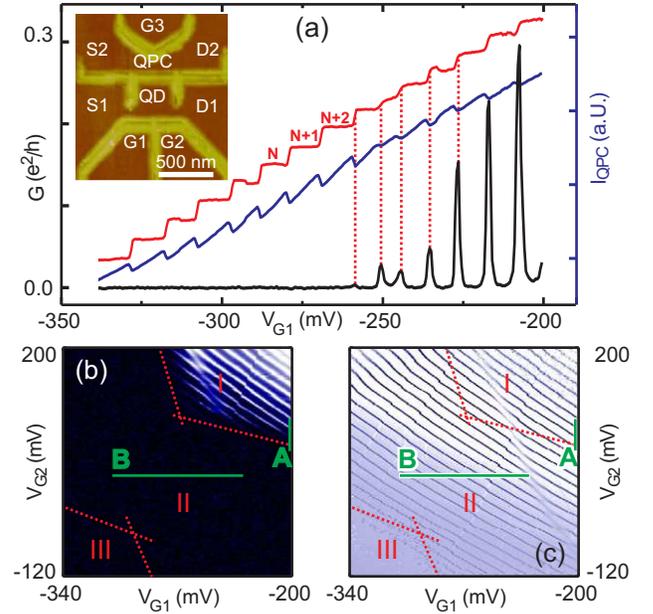}
\caption{(a) differential QD conductance $G$, QPC current
$I_{QPC}$ and dot charge as a function of $V_{G1}$. $G$ shows
Coulomb blockade peaks typical for single quantum dots. At each
peak a step in $I_{QPC}$ is visible due to capacitive coupling.
This still holds when the Coulomb peaks vanish. Inset: AFM image
of the device. (b) Charge diagram for $G$. Bright means high
differential conductance, dark means low. (c) Derivative
$\frac{dI_{QPC}}{dV_G}$ of QPC current corresponding to (b). Dark
means charging event, bright means constant charge. The color
encoding is used throughout this work.} \label{fig1}
\end{figure}

Our QD device is based on a GaAs/AlGaAs heterostructure with a
two-dimensional electron system (2DES) 34~nm below the surface.
The sheet density is $n=4.59\times 10^{15}$~m$^{-2}$, the mobility
is $\mu=64.3$~m$^2/$Vs. We use an atomic force microscope (AFM) to
write oxide lines on the sample surface by local anodic oxidation
(LAO) \cite{Keyser-00}. These oxide lines deplete the 2DES below
and thus create an electronic potential in the plane of the 2DES
forming the coupled QD/QPC system \cite{Fricke-05}. They are
visible as bright lines in the AFM image of our device [inset of
Fig.\ \ref{fig1}(a)]. QD and QPC are coupled individually to
source and drain leads ($S1$ and $D1$ for the QD, $S2$ and $D2$
for the QPC). Two side gates $G1$ and $G2$ control the tunneling
rates of the dot, a third gate $G3$ controls the transparency of
the QPC.

Our device is characterized in a $^3$He/$^4$He dilution
refrigerator at a base temperature of 40~mK. We measure the
differential conductance $G$ through the QD and additionally the
dc current $I_{QPC}$ through the QPC. Due to its close vicinity to
the QD, the QPC is capacitively coupled to the charge on the dot.
Thus it is extremely sensitive to changes of the dot charge. In
Fig.\ \ref{fig1}(a) $G$ and $I_{QPC}$ are plotted as a function of
the voltage applied to gate $G1$. $G$ shows Coulomb peaks typical
for single quantum dots in an opaque regime. Each peak corresponds
to a charging event of the dot. Thus two adjacent minima reflect a
charge difference of one electron. Correlated with $G$, the QPC
current $I_{QPC}$ shows a sawtooth-like behavior. Each electron
entering the dot changes the potential of the QPC, thus reducing
the current. The charge on the dot is extracted by measuring the
difference in gate voltage between the measured QPC current and
the smooth curve of an ideal QPC which is influenced only by the
gate voltages but not by charging events on the dot. The charge
reveals a steplike behavior [see upper curve in Fig.\
\ref{fig1}(a)] \cite{charge-profile}.

With decreasing gate voltage, the Coulomb peaks vanish due to
reduced tunneling rates. But still the dot is able to adjust its
charge and the QPC measurement shows the typical steps. The
accessible parameter space is extended. This is also visible in
the charge diagram in Figs.\ \ref{fig1}(b) and (c). Figure
\ref{fig1}(b) shows $G$ as a function of gate voltages $V_{G1}$
and $V_{G2}$. Figure \ref{fig1}(c) shows the derivative of the
corresponding QPC current. The measurement can be divided into
three regions depending on the transparency of the dot barriers.
Region I corresponds to the configuration where both barriers are
sufficiently permeable for transport measurements. This transport
region is bounded to the left and to the bottom marked with two
dotted lines. Here the barriers lose their ability to give a
measurable differential conductance (currents $>0.2$~pA). But
still they allow for charging events visible in the QPC
measurement in region II. This charging region is again bounded
towards lower voltages marked by two lines parallel to those lines
limiting region I. These new lines correspond to the
configuration, where the barriers decouple the dot from the leads
so that charging becomes impossible. Thus in region III no signal
is observed, neither in QD nor in QPC transport.

\begin{figure}
\includegraphics[scale=1]{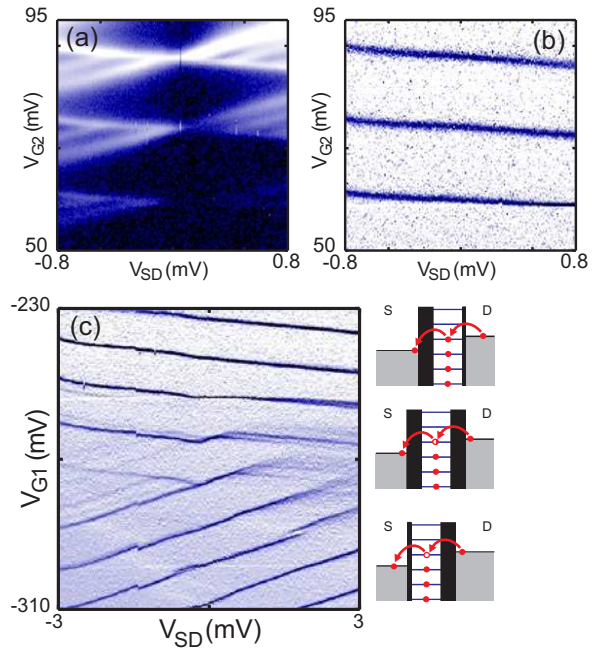}
\caption{Nonlinear measurements along line A  and line B (compare
Fig.\ \ref{fig1}). Along line A, full diamonds are visible in the
differential QD conductance (a) but only lines with negative
slopes in the QPC current (b) due to asymmetric tunneling rates.
Along line B, a transition from lines with positive slopes to
lines with negative slopes appears in the QPC current (c). In the
middle, a symmetric configuration is found where both slopes and
excited states are visible. The sketches indicate the
configurations of the tunneling barriers.} \label{fig2}
\end{figure}

In addition to the extension of the parameter space, we will show
that the QPC can also be used to investigate the symmetry of the
tunneling rates $\Gamma_S$ and $\Gamma_D$. For this purpose we
performed nonlinear measurements. Figures \ref{fig2}(a) and (b)
show a measurement along line A in Figs.\ \ref{fig1}(b) and (c).
$G$ and the derivative of $I_{QPC}$ are plotted as a function of
$V_{SD}$ and $V_{G2}$ [Figs. \ref{fig2}(a) and \ref{fig2}(b),
respectively]. The differential conductance shows the typical
Coulomb diamonds including excited dot states. Lines with positive
slope correspond to dot transitions due to resonances of QD states
with source (with chemical potential $\mu_S$) while lines with
negative slope correspond to resonances with drain ($\mu_D$).
Although $G$ shows full diamonds, the QPC measurement reveals only
the lines with negative slope. The lines with positive slope as
well as lines corresponding to transitions involving excited
states are not visible. Thus the charge on the dot is not altered
by these transitions. Only when a ground state (with a chemical
potential $\mu_N$) comes in resonance with drain ($\mu_N=\mu_D$) a
charging event is detected.

We find the same configuration in the upper part of Fig.\
\ref{fig2}(c) which is taken along line B in Figs.\ \ref{fig1}(b)
and (c). Again, only lines with negative slopes are visible. But
in contrast to Fig.\ \ref{fig2}(b) the configuration changes in
the lower part of Fig.\ \ref{fig2}(c). Here only lines with
positive slopes are visible. Thus here resonances of ground states
with source are detected ($\mu_N=\mu_S$). Only in the middle of
the figure both slopes are detected. Ground state transitions are
visible as well as transitions with excited states.

This is explained taking into account the ratio of the barriers'
tunneling rates. Two basic asymmetric configurations can be
distinguished [see sketches in Fig.\ \ref{fig2}(c)]. Either the
barrier on the source side is more opaque ($\Gamma_S<\Gamma_D$) or
the one on the drain side ($\Gamma_S>\Gamma_D$). If
$\Gamma_S<\Gamma_D$ the dot charge is governed by the drain's
chemical potential $\mu_D$. On average an $N$-electron ground
state with $\mu_N>\mu_D$ is empty even when the chemical potential
of the source enables electrons to tunnel on this state. The
electrons leave the dot through the drain much faster than new
electrons enter the dot from the source. On the other hand, if
$\mu_N<\mu_D$ the state is filled on average even if the source
potential enables electrons to leave the dot through the source.
Those electrons are immediately replaced with new electrons
injected from the drain. Thus the dot charge is altered only when
$\mu_N=\mu_D$ which gives a line with negative slope in the QPC
measurement. Excited states do not affect the dot's charge. Either
they are empty because their chemical potentials are above $\mu_D$
or their occupation does not change the total charge because the
dot was already filled with $N$ electrons due to the occupied
$N$-electron ground state. If the drain is more opaque
($\Gamma_S>\Gamma_D$) the situation is the opposite. Now only
lines with a positive slope are visible.

While Fig.\ \ref{fig2}(b) shows only the configuration with
$\Gamma_S<\Gamma_D$, the transition between both configurations is
visible in Fig.\ \ref{fig2}(c). In the upper part of the figure,
the lines with a negative slope reflect the configuration
$\Gamma_S<\Gamma_D$. The lower part features $\Gamma_S>\Gamma_D$.
In between, we find the transition where lines for both slopes
appear. In this configuration the tunneling rates of both barriers
are equal ($\Gamma_S=\Gamma_D$). Transitions are detected for
ground states as well as excited states at resonance with source
and with drain.

\begin{figure}
\includegraphics[scale=1]{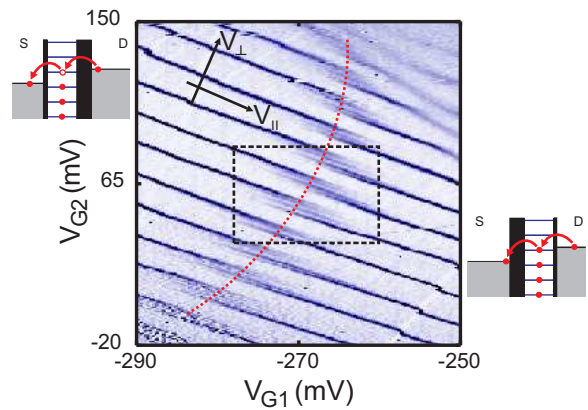}
\caption{Derivative of QPC current for a charge diagram similar to
Fig.\ \ref{fig1} but with $V_{SD}=1$~mV and $B=3.7$~T. On the left
and right single lines are visible which correspond to ground
state transitions. The lines on the left appear at higher voltages
(higher $V_\bot$) than those on the right. Thus on the left
(right) the drain (source) barrier is more opaque (see sketches).
In the middle a symmetry line is observed (marked dotted) where
both lines appear along with excited states.} \label{fig3}
\end{figure}

These measurements demonstrate that one can find symmetric
configurations by sweeping $V_{SD}$. This is still possible when a
perpendicular magnetic field is applied. A charge diagram similar
to Fig.\ \ref{fig1} (c) for $B=3.7$~T is shown in Fig.\
\ref{fig3}. The QPC current is recorded as a function of $V_{G1}$
and $V_{G2}$. The detection of symmetric configurations is now
realized with $V_{SD}$ set to 1~mV. When $V_{SD}>0$, resonances
with source appear at higher gate voltages than with $V_{SD}=0$
and resonances with drain at lower voltages. This is visible in
Fig.\ \ref{fig3}. The lines on the left appear at higher voltages
(higher $V_\bot$, see Fig.\ \ref{fig3}) than those on the right.
In the middle, both lines are visible with additional lines in
between. These additional lines correspond to excited states
revealing a symmetric configuration while the lines on the left
and right are those for ground state transitions due to resonance
with source or drain. Overall this measurement reveals a complete
line of symmetric barriers (marked as dotted line) that connects
region I with region III introduced in the context of Fig.\
\ref{fig1}. On the left we have $\Gamma_S>\Gamma_D$ and on the
right $\Gamma_S<\Gamma_D$.

\begin{figure}
\includegraphics[scale=1.1]{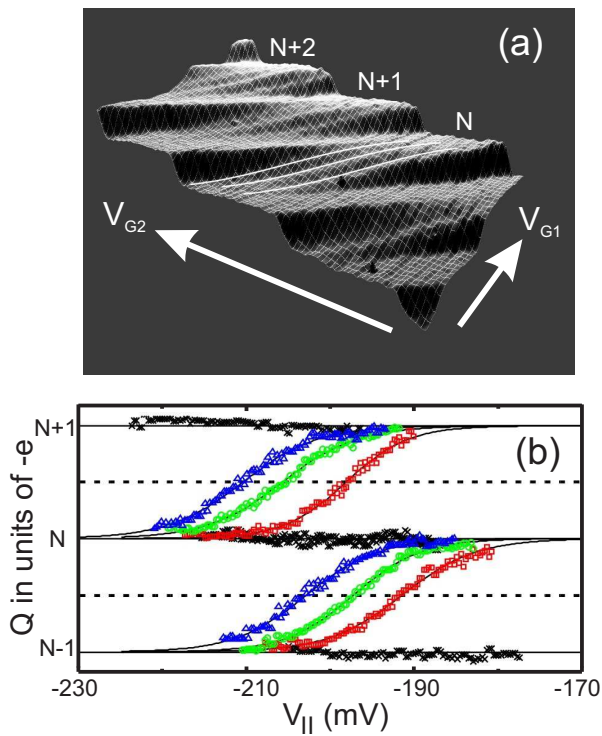}
\caption{(a) 3D plot of the QD charge for the region marked in
Fig.\ \ref{fig3}. The large steps correspond to the addition of
electrons. Several smaller steps marked by white lines reflect
resonances of ground and excited states near the symmetric
configuration. The charging for these resonances and for those at
$N \rightarrow N+1$ is studied in detail in (b) as a function of
$V_\|$.} \label{fig4}
\end{figure}

Now we will investigate the symmetric configuration in more
detail. For this purpose we take a closer look at the region
marked with a dashed rectangle in Fig.\ \ref{fig3}. The charge of
the dot for this region is shown in a 3D plot in Fig.\
\ref{fig4}(a). A few large steps that are equal in height are
clearly visible. They correspond to the addition of electrons to
the dot leading to electron numbers $N$, $N+1$, $N+2$ and so on.
Furthermore, each step features several smaller steps that
continuously lead from a smaller electron number to a larger one.
Three of them at the transition $N-1$ to $N$ are marked with white
lines. These features correspond to the occupation of ground and
excited states visible due to the transition $\Gamma_S>\Gamma_D
\rightarrow \Gamma_S<\Gamma_D$.

The charge along the marked lines and along the features at the
next large step is plotted in Fig.\ \ref{fig4}(b) as a function of
$V_\|$ which is the voltage parallel to the Coulomb blockade peaks
in a rotated coordinate system (see Fig.\ \ref{fig3}). The exact
position of symmetric coupling ($\Gamma_S=\Gamma_D$) is given for
half-integer electron numbers at the intersections with the dashed
lines in Fig.\ \ref{fig4} (b). From additional nonlinear
measurements we know that all features between constant electron
numbers appear due to resonances with drain. Therefore the lowest
lines [squares] corresponding to the occupation of a ground state
reflect the coupling characteristics of only this state, because
no other state is used as a transport channel. Thus the
configuration of equal tunneling rates is identical with the
ground state placed symmetrically between both leads. The other
lines [circles and triangles] reflect resonances of excited states
which contribute to the transport only together with the ground
state. Thus only the coupling characteristics of the total system
can be detected, but not the symmetry of each individual state.
This is confirmed by the observation that the symmetric
configuration shifts to lower voltages with the inclusion of
excited states as transport channels. If more than one channel is
in the transport window electrons have several ways to enter the
dot but still only one way to exit. Thus the mean occupation
increases even if the barriers remain unchanged. Consequently, the
symmetric configuration shifts to lower voltages. This effect is
visible at all large steps.

However there are differences between the transitions $N-1
\rightarrow N$ and $N \rightarrow N+1$. The positions of the lines
for excited states are not exactly the same. The line for the
first excited state at $N-1 \rightarrow N$ appears closer to the
line for the corresponding ground state than at $N \rightarrow
N+1$. This is not due to a larger energy difference between ground
and excited state. The difference in $V_\bot$ which corresponds to
the energy difference is even bigger for $N-1 \rightarrow N$ (as
can be seen in Fig.\ \ref{fig3}). The difference in $V_\|$ is
rather influenced by the spatial distribution of ground and
excited states. Thus for $N-1 \rightarrow N$ the excited state
seems to be shifted to the source lead compared to $N \rightarrow
N+1$. Thus the coupling symmetry is influenced not only by the
barriers which are controlled by the gate voltages. It is also
affected by the number of states contributing to transport and
their spacial distribution.

In summary, we performed transport measurements on a coupled
system containing a QD and a QPC. We used the QPC to detect
charging events on the dot. In this way the accessible parameter
space was extended. We used the charge detection to investigate
the ratio of the tunneling rates to source and drain leads of the
dot in dc measurements. We found a line of symmetry where both
tunneling rates are equal. Along this line ground as well as
excited states were detected. Their behavior concerning symmetric
tunneling rates was studied in detail. The influence of the
spatial distribution of ground and excited dot states was
investigated and a dependence on the number of transport channels
was found.
\\\\
This work has been supported by BMBF. We thank R. Winkler for
critically reading the manuscript.




\begin{thebibliography}{20}
\expandafter\ifx\csname natexlab\endcsname\relax\def\natexlab#1{#1}\fi
\expandafter\ifx\csname bibnamefont\endcsname\relax
  \def\bibnamefont#1{#1}\fi
\expandafter\ifx\csname bibfnamefont\endcsname\relax
  \def\bibfnamefont#1{#1}\fi
\expandafter\ifx\csname citenamefont\endcsname\relax
  \def\citenamefont#1{#1}\fi
\expandafter\ifx\csname url\endcsname\relax
  \def\url#1{\texttt{#1}}\fi
\expandafter\ifx\csname urlprefix\endcsname\relax\def\urlprefix{URL }\fi
\providecommand{\bibinfo}[2]{#2}
\providecommand{\eprint}[2][]{\url{#2}}

\bibitem[{\citenamefont{Kouwenhoven et~al.}(1997)\citenamefont{Kouwenhoven,
  Marcus, McEuen, Tarucha, Westerveld, and Wingreen}}]{Kouwenhoven-97}
\bibinfo{author}{\bibfnamefont{L.~P.} \bibnamefont{Kouwenhoven}},
  \bibinfo{author}{\bibfnamefont{C.~M.} \bibnamefont{Marcus}},
  \bibinfo{author}{\bibfnamefont{P.~L.} \bibnamefont{McEuen}},
  \bibinfo{author}{\bibfnamefont{S.}~\bibnamefont{Tarucha}},
  \bibinfo{author}{\bibfnamefont{R.~M.} \bibnamefont{Westerveld}},
  \bibnamefont{and} \bibinfo{author}{\bibfnamefont{N.~S.}
  \bibnamefont{Wingreen}}, in \emph{\bibinfo{booktitle}{Mesoscopic Electron
  Transport}}, edited by \bibinfo{editor}{\bibfnamefont{L.~L.}
  \bibnamefont{Sohn}}, \bibinfo{editor}{\bibfnamefont{L.~P.}
  \bibnamefont{Kouwenhoven}}, \bibnamefont{and}
  \bibinfo{editor}{\bibfnamefont{G.}~\bibnamefont{Sch\"o{}n}}
  (\bibinfo{publisher}{Kluwer}, \bibinfo{address}{Dordrecht},
  \bibinfo{year}{1997}), vol. \bibinfo{volume}{345} of
  \emph{\bibinfo{series}{Series E}}, pp. \bibinfo{pages}{105--214}.


  \bibitem[{\citenamefont{Field et~al.}(1993)\citenamefont{Field, Smith, Pepper, Richie, Frost, Jones, and Hasko}}]{Field-93}
\bibinfo{author}{\bibfnamefont{M.} \bibnamefont{Field}},
\bibinfo{author}{\bibfnamefont{C.~G.}~\bibnamefont{Smith}},
\bibinfo{author}{\bibfnamefont{M.}~\bibnamefont{Pepper}},
\bibinfo{author}{\bibfnamefont{D.~A.}~\bibnamefont{Ritchie}},
\bibinfo{author}{\bibfnamefont{J.~E.~F.}~\bibnamefont{Frost}},
\bibinfo{author}{\bibfnamefont{G.~A.~C.}~\bibnamefont{Jones}},
\bibnamefont{and} \bibinfo{author}{\bibfnamefont{D.~G.} \bibnamefont{Hasko}},
  \bibinfo{journal}{Phys. Rev. Lett.} \textbf{\bibinfo{volume}{70}},
  \bibinfo{pages}{1311} (\bibinfo{year}{1993}).

\bibitem[{\citenamefont{Elzerman et~al.}(2003)\citenamefont{Elzerman, Hanson, Greidanus, Willems van Beveren, De Franceschi, Vandersypen, Tarucha, and Kouwenhoven}}]{Elzerman-03}
\bibinfo{author}{\bibfnamefont{J.~M.} \bibnamefont{Elzerman}},
\bibinfo{author}{\bibfnamefont{R.}~\bibnamefont{Hanson}},
\bibinfo{author}{\bibfnamefont{J.~S.}~\bibnamefont{Greidenus}},
\bibinfo{author}{\bibfnamefont{L.~H.}~\bibnamefont{Willems van Beveren}},
\bibinfo{author}{\bibfnamefont{S.}~\bibnamefont{De Franceschi}},
\bibinfo{author}{\bibfnamefont{L.~M.~K.}~\bibnamefont{Vandersypen}},
\bibinfo{author}{\bibfnamefont{S.}~\bibnamefont{Tarucha}},
\bibnamefont{and} \bibinfo{author}{\bibfnamefont{L.~P.} \bibnamefont{Kouwenhoven}},
  \bibinfo{journal}{Phys. Rev. B} \textbf{\bibinfo{volume}{67}},
  \bibinfo{pages}{161308(R)} (\bibinfo{year}{2003}).

  \bibitem[{\citenamefont{Nemutudi et~al.}(2004)\citenamefont{Nemutudi, Kataoka, Ford, Appleyard, Pepper, Ritchie, and Jones}}]{Nemutudi-04}
\bibinfo{author}{\bibfnamefont{R.} \bibnamefont{Nemutudi}},
\bibinfo{author}{\bibfnamefont{M.}~\bibnamefont{Kataoka}},
\bibinfo{author}{\bibfnamefont{C.~J.~B.}~\bibnamefont{Ford}},
\bibinfo{author}{\bibfnamefont{N.~J.}~\bibnamefont{Appleyard}},
\bibinfo{author}{\bibfnamefont{M.}~\bibnamefont{Pepper}},
\bibinfo{author}{\bibfnamefont{D.~A.}~\bibnamefont{Ritchie}},
\bibnamefont{and} \bibinfo{author}{\bibfnamefont{G.~A.~C.} \bibnamefont{Jones}},
  \bibinfo{journal}{J. Appl. Phys.} \textbf{\bibinfo{volume}{95}},
  \bibinfo{pages}{2557} (\bibinfo{year}{2004}).

\bibitem[{\citenamefont{Schleser et~al.}(2004)\citenamefont{Schleser, Ruh, Ihn, Ensslin, Driscoll, and Gossard}}]{Schleser-APL04}
\bibinfo{author}{\bibfnamefont{R.} \bibnamefont{Schleser}},
\bibinfo{author}{\bibfnamefont{E.}~\bibnamefont{Ruh}},
\bibinfo{author}{\bibfnamefont{T.}~\bibnamefont{Ihn}},
\bibinfo{author}{\bibfnamefont{K.}~\bibnamefont{Ensslin}},
\bibinfo{author}{\bibfnamefont{D.~C.}~\bibnamefont{Driscoll}},
\bibnamefont{and} \bibinfo{author}{\bibfnamefont{A.~C.} \bibnamefont{Gossard}},
  \bibinfo{journal}{Appl. Phys. Lett.} \textbf{\bibinfo{volume}{85}},
  \bibinfo{pages}{2005} (\bibinfo{year}{2004}).

  \bibitem[{\citenamefont{Fricke et~al.}(2004)\citenamefont{Fricke, Rogge, Harke, Reinwald, Wegscheider, Hohls, and Haug}}]{Fricke-05}
\bibinfo{author}{\bibfnamefont{C.} \bibnamefont{Fricke}},
\bibinfo{author}{\bibfnamefont{M.~C.} \bibnamefont{Rogge}},
\bibinfo{author}{\bibfnamefont{B.}~\bibnamefont{Harke}},
\bibinfo{author}{\bibfnamefont{M.} \bibnamefont{Reinwald}},
\bibinfo{author}{\bibfnamefont{W.} \bibnamefont{Wegscheider}},
\bibinfo{author}{\bibfnamefont{F.}~\bibnamefont{Hohls}},
\bibnamefont{and} \bibinfo{author}{\bibfnamefont{R.~J.} \bibnamefont{Haug}},
  \bibinfo{journal}{cond-mat/0504777} (\bibinfo{year}{2005}).

\bibitem[{\citenamefont{Elzerman et~al.}(2004)\citenamefont{Elzerman, Hanson, Willems van Beveren, Witkamp, Vandersypen, and Kouwenhoven}}]{Elzerman-Nature04}
\bibinfo{author}{\bibfnamefont{J.~M.} \bibnamefont{Elzerman}},
\bibinfo{author}{\bibfnamefont{R.}~\bibnamefont{Hanson}},
\bibinfo{author}{\bibfnamefont{L.~H.}~\bibnamefont{Willems van Beveren}},
\bibinfo{author}{\bibfnamefont{B.}~\bibnamefont{Witkamp}},
\bibinfo{author}{\bibfnamefont{L.~M.~K.}~\bibnamefont{Vandersypen}},
\bibnamefont{and} \bibinfo{author}{\bibfnamefont{L.~P.} \bibnamefont{Kouwenhoven}},
  \bibinfo{journal}{Nature} \textbf{\bibinfo{volume}{430}},
  \bibinfo{pages}{431} (\bibinfo{year}{2004}).

\bibitem[{\citenamefont{Elzerman et~al.}(2004)\citenamefont{Elzerman, Hanson, Willems van Beveren, Vandersypen, and Kouwenhoven}}]{Elzerman-APL04}
\bibinfo{author}{\bibfnamefont{J.~M.} \bibnamefont{Elzerman}},
\bibinfo{author}{\bibfnamefont{R.}~\bibnamefont{Hanson}},
\bibinfo{author}{\bibfnamefont{L.~H.}~\bibnamefont{Willems van Beveren}},
\bibinfo{author}{\bibfnamefont{L.~M.~K.}~\bibnamefont{Vandersypen}},
\bibnamefont{and} \bibinfo{author}{\bibfnamefont{L.~P.} \bibnamefont{Kouwenhoven}},
  \bibinfo{journal}{Appl. Phys. Lett.} \textbf{\bibinfo{volume}{84}},
  \bibinfo{pages}{4617} (\bibinfo{year}{2004}).

\bibitem[{\citenamefont{Johnson et~al.}(2005)\citenamefont{Johnson, Marcus, Hanson, and Gossard}}]{Johnson-PRB05}
\bibinfo{author}{\bibfnamefont{A.~C.} \bibnamefont{Johnson}},
\bibinfo{author}{\bibfnamefont{C.~M.}~\bibnamefont{Marcus}},
\bibinfo{author}{\bibfnamefont{M.~P.}~\bibnamefont{Hanson}},
\bibnamefont{and} \bibinfo{author}{\bibfnamefont{A.~C.} \bibnamefont{Gossard}},
  \bibinfo{journal}{Phys. Rev. B} \textbf{\bibinfo{volume}{71}},
  \bibinfo{pages}{115333} (\bibinfo{year}{2005}).

  \bibitem[{\citenamefont{Cavaliere et~al.}(2004)\citenamefont{Cavaliere, Braggio, Stockburger, Sassetti, and Kramer}}]{Cavaliere-04}
\bibinfo{author}{\bibfnamefont{F.} \bibnamefont{Cavaliere}},
\bibinfo{author}{\bibfnamefont{A.}~\bibnamefont{Braggio}},
\bibinfo{author}{\bibfnamefont{J.~T.}~\bibnamefont{Stockburger}},
\bibinfo{author}{\bibfnamefont{M.}~\bibnamefont{Sassetti}},
\bibnamefont{and} \bibinfo{author}{\bibfnamefont{B.} \bibnamefont{Kramer}},
  \bibinfo{journal}{Phys. Rev. Lett.} \textbf{\bibinfo{volume}{93}},
  \bibinfo{pages}{036803} (\bibinfo{year}{2004}).

\bibitem[{\citenamefont{Keyser et~al.}(2000)\citenamefont{Keyser, Schumacher,
  Zeitler, Haug, and Eberl}}]{Keyser-00}
\bibinfo{author}{\bibfnamefont{U.~F.} \bibnamefont{Keyser}},
  \bibinfo{author}{\bibfnamefont{H.~W.} \bibnamefont{Schumacher}},
  \bibinfo{author}{\bibfnamefont{U.}~\bibnamefont{Zeitler}},
  \bibinfo{author}{\bibfnamefont{R.~J.} \bibnamefont{Haug}}, \bibnamefont{and}
  \bibinfo{author}{\bibfnamefont{K.}~\bibnamefont{Eberl}},
  \bibinfo{journal}{Appl. Phys. Lett.} \textbf{\bibinfo{volume}{76}},
  \bibinfo{pages}{457} (\bibinfo{year}{2000}).

\bibitem[12]{charge-profile}With increasing differential conductance the QPC signal gets smoother. In addition, we observe a broadening and overlapping of two Coulomb peaks at
$V_{G1}=-250$~mV leading to a smooth charging profile. This is
ascribed to a local impurity in one barrier developing a single
molecule-like state also visible in Figs.\ \ref{fig1}(b) and (c).
\end{thebibliography}



\end{document}